%% file: main.tex
\documentclass[acmsmall,nonacm]{acmart}

\AtBeginDocument{%
  \providecommand\BibTeX{{%
    \normalfont B\kern-0.5em{\scshape i\kern-0.25em b}\kern-0.8em\TeX}}}

\setcopyright{acmcopyright}
\copyrightyear{2018}
\acmYear{2018}
\acmDOI{10.1145/1122445.1122456}

\acmConference[Woodstock '18]{Woodstock '18: ACM Symposium on Neural
  Gaze Detection}{June 03--05, 2018}{Woodstock, NY}
\acmBooktitle{Woodstock '18: ACM Symposium on Neural Gaze Detection,
  June 03--05, 2018, Woodstock, NY}
\acmPrice{15.00}
\acmISBN{978-1-4503-XXXX-X/18/06}

\begin{document}

\title{Descriptive AI Ethics: Collecting and Understanding the Public Opinion}



\author{Gabriel Lima}
\email{gabriel.lima@kaist.ac.kr}
\affiliation{%
  \institution{School of Computing, KAIST}
}
\affiliation{%
  \institution{Data Science Group, IBS}
  \country{Republic of Korea}
}

\author{Meeyoung Cha}
\email{mcha@ibs.re.kr}
\affiliation{%
  \institution{Data Science Group, IBS}
}
\affiliation{%
  \institution{School of Computing, KAIST}
  \country{Republic of Korea}
}

\renewcommand{\shortauthors}{Lima et al.}

\begin{abstract}
\input{content/0abstract}
\end{abstract}




\maketitle

\section{Introduction}
\input{content/1responsibility}

\section{Importance of the Public Opinion}
\input{content/2public}

\section{Normative and Data-Driven AI Ethics}
\input{content/3proposal}

\bibliographystyle{ACM-Reference-Format}
\bibliography{sample-base}










\end{document}

%% file: content/0abstract.tex
There is a growing need for data-driven research efforts on how the public perceives the ethical, moral, and legal issues of autonomous AI systems. The current debate on the responsibility gap posed by these systems is one such example. This work proposes a mixed AI ethics model that allows normative and descriptive research to complement each other, by aiding scholarly discussion with data gathered from the public. We discuss its implications on bridging the gap between optimistic and pessimistic views towards AI systems' deployment.

%% file: content/1responsibility.tex
In light of the significant changes artificial intelligence (AI) systems bring to society, many scholars discuss whether and how their influence can be positive and negative~\cite{floridi2018ai4people}. As we start to encounter AI systems in various morally and legally salient environments, some have begun to explore how the current responsibility ascription practices might be adapted to meet such new technologies~\cite{reddy2019beyond,floridi2016faultless}.

A critical viewpoint today is that autonomous and self-learning AI systems pose a so-called responsibility gap~\cite{matthias2004responsibility}. These systems' autonomy challenges human control over them~\cite{coeckelbergh2019artificial}, while their adaptability leads to unpredictability. Hence, it might infeasible to trace back responsibility to a specific entity if these systems cause any harm.  

Considering responsibility practices as the adoption of certain attitudes towards an agent~\cite{wallace1994responsibility}, scholarly work has also posed the question of whether AI systems are appropriate subjects of such practices~\cite{sparrow2007killer,mulligan2017revenge,danaher2016robots} --- e.g., they might ``have a body to kick,'' yet they ``have no soul to damn''~\cite{asaro201111}.

The dilemma of the responsibility gap is not restricted to moral domains but extends to legal practices~\cite{asaro2016liability,beck2016problem}. Current legal institutions are not capable of dealing with these questions~\cite{schirmer}. The nature of self-learning algorithms challenges the proximate causal connection between these systems' designers and possible damages~\cite{mulligan2017revenge}. There makes no sense to hold these systems themselves liable for their actions if they cannot remedy those harmed or learn from mistakes. The punishment of AI systems is hence an open and controversial question~\cite{asaro2007robots}.

We focus on the limitation of current normative efforts that address the responsibility gap. They discuss how AI systems \emph{could} and \emph{should} be embedded into our responsibility practices. Viewing responsibility as a relational concept in which one holds a wrongdoer responsible for a specific action or consequence~\cite{coeckelbergh2019artificial}, or in which an agent is responsible to a patient~\cite{duff2004responsible}, existing research disregards the opinion of the individuals involved in this practice. There lacks empirical work understanding how the general public, those who might suffer damages by AI systems, perceives these gaps and believes how these issues could be solved. Some studies exist in the domain of autonomous vehicles~\cite{awad2020drivers,li2016trolley}, yet the broader understanding of how people assign responsibility, blame, and punishment for the actions of autonomous systems is missing; AI is deployed in diverse forms and environments, and little work~\cite{lima2020punishing,van2019robots} has addressed other morally salient situations. 
 
The ethics of AI have been gaining much traction in the last years~\cite{tsamadosethics}. Many of the issues raised by the deployment of algorithms have been extensively debated in academia and industry. There have been attempts to understand how people perceive AI systems' ethical issues, such as bias/fairness~\cite{grgic2018human,saxena2019fairness} and privacy~\cite{youn2008gender}. However, the public opinion is yet to be captured and understood at a larger scale. In the following sections, we defend that the public opinion is precious and indispensable when discussing AI ethics, with a focus on responsibility gaps, and propose how normative work on the topic should embed data-driven research in their discussion so that those who do and do not believe these gaps can be bridged can find common ground.

%% file: content/2public.tex
Although normative work on the responsibility gaps might seem to provide a ``correct'' answer, these issues should be discussed publicly, especially when so much is at stake~\cite{coeckelbergh2019artificial}. The general public is an essential stakeholder in the deployment of these systems, and their opinion should be weighed against all other entities in society so that the ``algorithmic social contract'' can be crafted and maintained~\cite{rahwan2018society}. ``The ascription of moral and legal responsibility is always mediated through the folk-psychological understanding of agency,'' i.e., the assignment of responsibility depends on whether the general public perceives AI systems as agents~\cite{brozek2017legal}.

Public organizations have called for more participatory development of AI~\cite{unesco}. The AI4People initiative~\cite{floridi2018ai4people}, composed of leading scholars in AI ethics, has recommended the elicitation of the public opinion and understanding of AI and its applications. Moreover, the same proposal defends that existing legal institutions' capacity to compensate those harmed by AI should be assessed based on majority-agreed foundations. In the field of big data, scholars have also proposed to adopt ``participatory design'' with a focus on democratic participation in the development and deployment of algorithms~\cite{whitman2018potential}.

The public might also have conflicting views that must be addressed by future policymaking. For instance, empirical findings indicate that people assign blame~\cite{awad2020drivers} and responsibility~\cite{li2016trolley} to automated systems, although to a lesser extent than their human counterparts. They also desire to punish these systems, although they are aware that doing so is not possible or successful~\cite{lima2020punishing}, following the scholarly work raising doubt on the viability of AI systems' punishment~\cite{sparrow2007killer,bryson2017and}. These contradictions should be solved proactively rather than reactively so that these systems and policymaking reflect public expectations. 

Scholars have defended that designers should promote public discussion regarding the accountability for AI systems~\cite{ananny2018seeing}. However, corporations, which are often the designers and manufacturers of these systems, have incentives to ignore public input and shift regulation. These interest groups have an extreme influence on policymaking~\cite{domhoff1998rules}. The designers of AI systems might utilize their creations as liability shields~\cite{bryson2017and} and for agency laundering, by transferring deserved blame to the system~\cite{rubel2019agency}. Corporations might choose to implement superficial measures to appear more ethically aligned than they are~\cite{floridi2019translating}. Previous research has argued that corporate (social) responsibility has failed in current oligopolistic business practices~\cite{devinney}. This conflict of interest might indicate that academic and public organizations should take the lead in understanding how to embed public values into AI and their regulation.

There lacks empirical work on the public perception of AI ethics. Nonetheless, the few studies addressing the public perception of algorithmic fairness exemplify the importance of the public opinion on their design and regulation. Public preference toward fairness metrics is dependent on the environment where these systems are deployed~\cite{saxena2019fairness}. There exists much disagreement on which features are fair to use in AI-assisted bail decisions~\cite{grgic2018human}. Those most affected by AI systems must have a say in how they are regulated and designed so that their social impact can be improved~\cite{wong2019democratizing} by embedding the general public's values and opinions~\cite{rahwan2018society,whitman2018potential}.

%% file: content/3proposal.tex
\begin{figure}[t!]
  \centering
  \includegraphics[width=.4\linewidth]{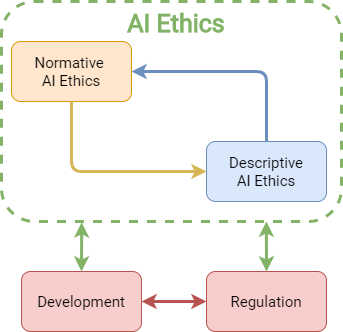}
  \caption{Model for combining data-driven research on AI ethics with its normative work on the development and regulation of AI systems.}
  \label{fig:model}
\end{figure}

As we have demonstrated above, understanding public opinion is vital to the successful and beneficial deployment of autonomous systems. Responsibility attribution is not an exception, as its practices are dependent on how the general public perceives the agents and patients of such interaction~\cite{brozek2017legal}. To obtain and make sense of people's opinions on the responsibility gap, we propose that ``interactive and discursive fora and processes'' that include all relevant stakeholders, including the general society, are necessary~\cite{buhmann2019managing}. 

A previous attempt to propose a method to crowdsource the public opinion on the moral dilemmas has proposed adding the "society to the loop."~\cite{awad2020crowdsourcing} However, much criticism has been given to its focus on \emph{dilemmas}, as they are not necessarily representative of most of the situations where AI is expected to influence society. It might not be possible to transfer the ethics of moral dilemmas, such as the trolley problem~\cite{nyholm2016ethics,mirnig2019trolled}, to the ethics of AI systems.

Previous work has also proposed to mathematically model moral judgments so that people's choices can be translated into policy statements~\cite{awad2020approach}. Although extremely valuable, it does not address the multiplicity of choices that an AI system might have when acting, or the plurality of judgments that the public might make. Modeling complex moral judgment mathematically as objective truths might also disregard society's social and cultural aspects, possibly re-enforcing harmful discrimination and societal problems~\cite{tsamadosethics}. Both quantitative and qualitative research methods should be applied in this domain so that the public opinion's intricacy on these issues can be successfully captured and analyzed.


We propose that the normative view on AI ethics, particularly the one dealing with the ascription of responsibility for automated systems' actions, should be augmented by a data-driven research agenda. Borrowing the concept from descriptive ethics, we advance a similar approach: descriptive AI ethics. We defend that an iterative model, in which empirical and normative research influence each other, is critical for the development and regulation of future AI systems (see Figure~\ref{fig:model}). 

Philosophical research on the responsibility gaps should raise questions and hypotheses that the descriptive approach to AI ethics should answer and test. With the public opinion on the topic provided by the data-driven approach, the normative agenda can deepen its analysis on how AI systems' development and regulation should move forward, making such results interpretable to non-experts and policymakers. Such findings should also set directions for the development of future systems. Those who lead AI development also play an important role in how these systems are regulated through various manners, such as lobbying and self-regulation.

We defend that such a model can contribute to bridging the gap between those who pose the responsibility gap can be solved~\cite{hanson2009beyond,nyholm2018attributing} and those more pessimistic about a solution~\cite{asaro_2012,danaher2016robots}. By aiding normative ethics research with data gathered from the public, more pragmatic approaches to the ethical issues of AI systems can be developed so that they are not only aligned with our values~\cite{zhu2018value,friedman1996value}, but also our solutions to their legal and moral questions.

%% file: main.bbl

\begin{thebibliography}{44}


\ifx \showCODEN    \undefined \def \showCODEN     #1{\unskip}     \fi
\ifx \showDOI      \undefined \def \showDOI       #1{#1}\fi
\ifx \showISBNx    \undefined \def \showISBNx     #1{\unskip}     \fi
\ifx \showISBNxiii \undefined \def \showISBNxiii  #1{\unskip}     \fi
\ifx \showISSN     \undefined \def \showISSN      #1{\unskip}     \fi
\ifx \showLCCN     \undefined \def \showLCCN      #1{\unskip}     \fi
\ifx \shownote     \undefined \def \shownote      #1{#1}          \fi
\ifx \showarticletitle \undefined \def \showarticletitle #1{#1}   \fi
\ifx \showURL      \undefined \def \showURL       {\relax}        \fi
\providecommand\bibfield[2]{#2}
\providecommand\bibinfo[2]{#2}
\providecommand\natexlab[1]{#1}
\providecommand\showeprint[2][]{arXiv:#2}

\bibitem[\protect\citeauthoryear{Ananny and Crawford}{Ananny and
  Crawford}{2018}]%
        {ananny2018seeing}
\bibfield{author}{\bibinfo{person}{Mike Ananny} {and} \bibinfo{person}{Kate
  Crawford}.} \bibinfo{year}{2018}\natexlab{}.
\newblock \showarticletitle{Seeing without knowing: Limitations of the
  transparency ideal and its application to algorithmic accountability}.
\newblock \bibinfo{journal}{\emph{New Media \& Society}} \bibinfo{volume}{20},
  \bibinfo{number}{3} (\bibinfo{year}{2018}), \bibinfo{pages}{973--989}.
\newblock


\bibitem[\protect\citeauthoryear{Asaro}{Asaro}{2012}]%
        {asaro_2012}
\bibfield{author}{\bibinfo{person}{Peter Asaro}.}
  \bibinfo{year}{2012}\natexlab{}.
\newblock \showarticletitle{On banning autonomous weapon systems: human rights,
  automation, and the dehumanization of lethal decision-making}.
\newblock \bibinfo{journal}{\emph{International Review of the Red Cross}}
  \bibinfo{volume}{94}, \bibinfo{number}{886} (\bibinfo{year}{2012}),
  \bibinfo{pages}{687–709}.
\newblock
\urldef\tempurl%
\url{https://doi.org/10.1017/S1816383112000768}
\showDOI{\tempurl}


\bibitem[\protect\citeauthoryear{Asaro}{Asaro}{2007}]%
        {asaro2007robots}
\bibfield{author}{\bibinfo{person}{Peter~M Asaro}.}
  \bibinfo{year}{2007}\natexlab{}.
\newblock \showarticletitle{Robots and responsibility from a legal
  perspective}.
\newblock \bibinfo{journal}{\emph{Proc. IEEE}} \bibinfo{volume}{4},
  \bibinfo{number}{14} (\bibinfo{year}{2007}), \bibinfo{pages}{20--24}.
\newblock


\bibitem[\protect\citeauthoryear{Asaro}{Asaro}{2011}]%
        {asaro201111}
\bibfield{author}{\bibinfo{person}{Peter~M Asaro}.}
  \bibinfo{year}{2011}\natexlab{}.
\newblock \showarticletitle{11 A Body to Kick, but Still No Soul to Damn: Legal
  Perspectives on Robotics}.
\newblock \bibinfo{journal}{\emph{Robot Ethics: The Ethical and Social
  Implications of Robotics}} (\bibinfo{year}{2011}), \bibinfo{pages}{169}.
\newblock


\bibitem[\protect\citeauthoryear{Asaro}{Asaro}{2016}]%
        {asaro2016liability}
\bibfield{author}{\bibinfo{person}{Peter~M Asaro}.}
  \bibinfo{year}{2016}\natexlab{}.
\newblock \showarticletitle{The Liability Problem for Autonomous Artificial
  Agents.}. In \bibinfo{booktitle}{\emph{AAAI Spring Symposia}}.
\newblock


\bibitem[\protect\citeauthoryear{Awad, Anderson, Anderson, and Liao}{Awad
  et~al\mbox{.}}{2020a}]%
        {awad2020approach}
\bibfield{author}{\bibinfo{person}{Edmond Awad}, \bibinfo{person}{Michael
  Anderson}, \bibinfo{person}{Susan~Leigh Anderson}, {and}
  \bibinfo{person}{Beishui Liao}.} \bibinfo{year}{2020}\natexlab{a}.
\newblock \showarticletitle{An approach for combining ethical principles with
  public opinion to guide public policy}.
\newblock \bibinfo{journal}{\emph{Artificial Intelligence}}
  (\bibinfo{year}{2020}), \bibinfo{pages}{103349}.
\newblock


\bibitem[\protect\citeauthoryear{Awad, Dsouza, Bonnefon, Shariff, and
  Rahwan}{Awad et~al\mbox{.}}{2020b}]%
        {awad2020crowdsourcing}
\bibfield{author}{\bibinfo{person}{Edmond Awad}, \bibinfo{person}{Sohan
  Dsouza}, \bibinfo{person}{Jean-Fran{\c{c}}ois Bonnefon},
  \bibinfo{person}{Azim Shariff}, {and} \bibinfo{person}{Iyad Rahwan}.}
  \bibinfo{year}{2020}\natexlab{b}.
\newblock \showarticletitle{Crowdsourcing moral machines}.
\newblock \bibinfo{journal}{\emph{Commun. ACM}} \bibinfo{volume}{63},
  \bibinfo{number}{3} (\bibinfo{year}{2020}), \bibinfo{pages}{48--55}.
\newblock


\bibitem[\protect\citeauthoryear{Awad, Levine, Kleiman-Weiner, Dsouza,
  Tenenbaum, Shariff, Bonnefon, and Rahwan}{Awad et~al\mbox{.}}{2020c}]%
        {awad2020drivers}
\bibfield{author}{\bibinfo{person}{Edmond Awad}, \bibinfo{person}{Sydney
  Levine}, \bibinfo{person}{Max Kleiman-Weiner}, \bibinfo{person}{Sohan
  Dsouza}, \bibinfo{person}{Joshua~B Tenenbaum}, \bibinfo{person}{Azim
  Shariff}, \bibinfo{person}{Jean-Fran{\c{c}}ois Bonnefon}, {and}
  \bibinfo{person}{Iyad Rahwan}.} \bibinfo{year}{2020}\natexlab{c}.
\newblock \showarticletitle{Drivers are blamed more than their automated cars
  when both make mistakes}.
\newblock \bibinfo{journal}{\emph{Nature Human Behaviour}} \bibinfo{volume}{4},
  \bibinfo{number}{2} (\bibinfo{year}{2020}), \bibinfo{pages}{134--143}.
\newblock


\bibitem[\protect\citeauthoryear{Beck}{Beck}{2016}]%
        {beck2016problem}
\bibfield{author}{\bibinfo{person}{Susanne Beck}.}
  \bibinfo{year}{2016}\natexlab{}.
\newblock \showarticletitle{The problem of ascribing legal responsibility in
  the case of robotics}.
\newblock \bibinfo{journal}{\emph{AI \& Society}} \bibinfo{volume}{31},
  \bibinfo{number}{4} (\bibinfo{year}{2016}), \bibinfo{pages}{473--481}.
\newblock


\bibitem[\protect\citeauthoryear{Bro{\.z}ek and Jakubiec}{Bro{\.z}ek and
  Jakubiec}{2017}]%
        {brozek2017legal}
\bibfield{author}{\bibinfo{person}{Bartosz Bro{\.z}ek} {and}
  \bibinfo{person}{Marek Jakubiec}.} \bibinfo{year}{2017}\natexlab{}.
\newblock \showarticletitle{On the legal responsibility of autonomous
  machines}.
\newblock \bibinfo{journal}{\emph{Artificial Intelligence and Law}}
  \bibinfo{volume}{25}, \bibinfo{number}{3} (\bibinfo{year}{2017}),
  \bibinfo{pages}{293--304}.
\newblock


\bibitem[\protect\citeauthoryear{Bryson, Diamantis, and Grant}{Bryson
  et~al\mbox{.}}{2017}]%
        {bryson2017and}
\bibfield{author}{\bibinfo{person}{Joanna~J Bryson},
  \bibinfo{person}{Mihailis~E Diamantis}, {and} \bibinfo{person}{Thomas~D
  Grant}.} \bibinfo{year}{2017}\natexlab{}.
\newblock \showarticletitle{Of, for, and by the people: the legal lacuna of
  synthetic persons}.
\newblock \bibinfo{journal}{\emph{Artificial Intelligence and Law}}
  \bibinfo{volume}{25}, \bibinfo{number}{3} (\bibinfo{year}{2017}),
  \bibinfo{pages}{273--291}.
\newblock


\bibitem[\protect\citeauthoryear{Buhmann, Pa{\ss}mann, and Fieseler}{Buhmann
  et~al\mbox{.}}{2019}]%
        {buhmann2019managing}
\bibfield{author}{\bibinfo{person}{Alexander Buhmann},
  \bibinfo{person}{Johannes Pa{\ss}mann}, {and} \bibinfo{person}{Christian
  Fieseler}.} \bibinfo{year}{2019}\natexlab{}.
\newblock \showarticletitle{Managing algorithmic accountability: Balancing
  reputational concerns, engagement strategies, and the potential of rational
  discourse}.
\newblock \bibinfo{journal}{\emph{Journal of Business Ethics}}
  (\bibinfo{year}{2019}), \bibinfo{pages}{1--16}.
\newblock


\bibitem[\protect\citeauthoryear{Coeckelbergh}{Coeckelbergh}{2019}]%
        {coeckelbergh2019artificial}
\bibfield{author}{\bibinfo{person}{Mark Coeckelbergh}.}
  \bibinfo{year}{2019}\natexlab{}.
\newblock \showarticletitle{Artificial intelligence, responsibility
  attribution, and a relational justification of explainability}.
\newblock \bibinfo{journal}{\emph{Science and Engineering Ethics}}
  (\bibinfo{year}{2019}), \bibinfo{pages}{1--18}.
\newblock


\bibitem[\protect\citeauthoryear{COMEST}{COMEST}{2017}]%
        {unesco}
\bibfield{author}{\bibinfo{person}{COMEST}.} \bibinfo{year}{2017}\natexlab{}.
\newblock \showarticletitle{Report of COMEST on Robotics Ethics}.
\newblock  (\bibinfo{year}{2017}).
\newblock
\newblock
\shownote{\url{https://unesdoc.unesco.org/ark:/48223/pf0000253952}. Accessed 16
  November 2019.}


\bibitem[\protect\citeauthoryear{Danaher}{Danaher}{2016}]%
        {danaher2016robots}
\bibfield{author}{\bibinfo{person}{John Danaher}.}
  \bibinfo{year}{2016}\natexlab{}.
\newblock \showarticletitle{Robots, law and the retribution gap}.
\newblock \bibinfo{journal}{\emph{Ethics and Information Technology}}
  \bibinfo{volume}{18}, \bibinfo{number}{4} (\bibinfo{year}{2016}),
  \bibinfo{pages}{299--309}.
\newblock


\bibitem[\protect\citeauthoryear{Devinney}{Devinney}{2009}]%
        {devinney}
\bibfield{author}{\bibinfo{person}{Timothy Devinney}.}
  \bibinfo{year}{2009}\natexlab{}.
\newblock \showarticletitle{Is the Socially Responsible Corporation a Myth? The
  Good, Bad and Ugly of Corporate Social Responsibility}.
\newblock \bibinfo{journal}{\emph{Academy of Management Perspectives}}
  \bibinfo{volume}{23} (\bibinfo{date}{03} \bibinfo{year}{2009}).
\newblock
\urldef\tempurl%
\url{https://doi.org/10.5465/AMP.2009.39985540}
\showDOI{\tempurl}


\bibitem[\protect\citeauthoryear{Domhoff}{Domhoff}{1998}]%
        {domhoff1998rules}
\bibfield{author}{\bibinfo{person}{G~William Domhoff}.}
  \bibinfo{year}{1998}\natexlab{}.
\newblock \bibinfo{booktitle}{\emph{Who rules America?: power and politics in
  the year 2000}}.
\newblock \bibinfo{publisher}{McGraw-Hill Humanities, Social Sciences \& World
  Languages}.
\newblock


\bibitem[\protect\citeauthoryear{Duff}{Duff}{2004}]%
        {duff2004responsible}
\bibfield{author}{\bibinfo{person}{Robin~Anthony Duff}.}
  \bibinfo{year}{2004}\natexlab{}.
\newblock \showarticletitle{Who is responsible, for what, to whom}.
\newblock \bibinfo{journal}{\emph{Ohio St. J. Crim. L.}}  \bibinfo{volume}{2}
  (\bibinfo{year}{2004}), \bibinfo{pages}{441}.
\newblock


\bibitem[\protect\citeauthoryear{Floridi}{Floridi}{2016}]%
        {floridi2016faultless}
\bibfield{author}{\bibinfo{person}{Luciano Floridi}.}
  \bibinfo{year}{2016}\natexlab{}.
\newblock \showarticletitle{Faultless responsibility: on the nature and
  allocation of moral responsibility for distributed moral actions}.
\newblock \bibinfo{journal}{\emph{Philosophical Transactions of the Royal
  Society A: Mathematical, Physical and Engineering Sciences}}
  \bibinfo{volume}{374}, \bibinfo{number}{2083} (\bibinfo{year}{2016}),
  \bibinfo{pages}{20160112}.
\newblock


\bibitem[\protect\citeauthoryear{Floridi}{Floridi}{2019}]%
        {floridi2019translating}
\bibfield{author}{\bibinfo{person}{Luciano Floridi}.}
  \bibinfo{year}{2019}\natexlab{}.
\newblock \showarticletitle{Translating principles into practices of digital
  ethics: Five risks of being unethical}.
\newblock \bibinfo{journal}{\emph{Philosophy \& Technology}}
  \bibinfo{volume}{32}, \bibinfo{number}{2} (\bibinfo{year}{2019}),
  \bibinfo{pages}{185--193}.
\newblock


\bibitem[\protect\citeauthoryear{Floridi, Cowls, Beltrametti, Chatila,
  Chazerand, Dignum, Luetge, Madelin, Pagallo, Rossi, et~al\mbox{.}}{Floridi
  et~al\mbox{.}}{2018}]%
        {floridi2018ai4people}
\bibfield{author}{\bibinfo{person}{Luciano Floridi}, \bibinfo{person}{Josh
  Cowls}, \bibinfo{person}{Monica Beltrametti}, \bibinfo{person}{Raja Chatila},
  \bibinfo{person}{Patrice Chazerand}, \bibinfo{person}{Virginia Dignum},
  \bibinfo{person}{Christoph Luetge}, \bibinfo{person}{Robert Madelin},
  \bibinfo{person}{Ugo Pagallo}, \bibinfo{person}{Francesca Rossi},
  {et~al\mbox{.}}} \bibinfo{year}{2018}\natexlab{}.
\newblock \showarticletitle{AI4People—an ethical framework for a good AI
  society: opportunities, risks, principles, and recommendations}.
\newblock \bibinfo{journal}{\emph{Minds and Machines}} \bibinfo{volume}{28},
  \bibinfo{number}{4} (\bibinfo{year}{2018}), \bibinfo{pages}{689--707}.
\newblock


\bibitem[\protect\citeauthoryear{Friedman}{Friedman}{1996}]%
        {friedman1996value}
\bibfield{author}{\bibinfo{person}{Batya Friedman}.}
  \bibinfo{year}{1996}\natexlab{}.
\newblock \showarticletitle{Value-Sensitive Design}.
\newblock \bibinfo{journal}{\emph{Interactions}} \bibinfo{volume}{3},
  \bibinfo{number}{6} (\bibinfo{date}{Dec.} \bibinfo{year}{1996}),
  \bibinfo{pages}{16–23}.
\newblock
\urldef\tempurl%
\url{https://doi.org/10.1145/242485.242493}
\showDOI{\tempurl}


\bibitem[\protect\citeauthoryear{Grgic-Hlaca, Redmiles, Gummadi, and
  Weller}{Grgic-Hlaca et~al\mbox{.}}{2018}]%
        {grgic2018human}
\bibfield{author}{\bibinfo{person}{Nina Grgic-Hlaca}, \bibinfo{person}{Elissa~M
  Redmiles}, \bibinfo{person}{Krishna~P Gummadi}, {and} \bibinfo{person}{Adrian
  Weller}.} \bibinfo{year}{2018}\natexlab{}.
\newblock \showarticletitle{Human perceptions of fairness in algorithmic
  decision making: A case study of criminal risk prediction}. In
  \bibinfo{booktitle}{\emph{Proc. of the World Wide Web Conference}}.
  \bibinfo{pages}{903--912}.
\newblock


\bibitem[\protect\citeauthoryear{Hanson}{Hanson}{2009}]%
        {hanson2009beyond}
\bibfield{author}{\bibinfo{person}{F~Allan Hanson}.}
  \bibinfo{year}{2009}\natexlab{}.
\newblock \showarticletitle{Beyond the skin bag: On the moral responsibility of
  extended agencies}.
\newblock \bibinfo{journal}{\emph{Ethics and Information Technology}}
  \bibinfo{volume}{11}, \bibinfo{number}{1} (\bibinfo{year}{2009}),
  \bibinfo{pages}{91--99}.
\newblock


\bibitem[\protect\citeauthoryear{Li, Zhao, Cho, Ju, and Malle}{Li
  et~al\mbox{.}}{2016}]%
        {li2016trolley}
\bibfield{author}{\bibinfo{person}{Jamy Li}, \bibinfo{person}{Xuan Zhao},
  \bibinfo{person}{Mu-Jung Cho}, \bibinfo{person}{Wendy Ju}, {and}
  \bibinfo{person}{Bertram~F Malle}.} \bibinfo{year}{2016}\natexlab{}.
\newblock \bibinfo{booktitle}{\emph{From trolley to autonomous vehicle:
  Perceptions of responsibility and moral norms in traffic accidents with
  self-driving cars}}.
\newblock \bibinfo{type}{{T}echnical {R}eport}. \bibinfo{institution}{SAE
  Technical Paper}.
\newblock


\bibitem[\protect\citeauthoryear{Lima, Jeon, Cha, and Park}{Lima
  et~al\mbox{.}}{2020}]%
        {lima2020punishing}
\bibfield{author}{\bibinfo{person}{Gabriel Lima}, \bibinfo{person}{Chihyung
  Jeon}, \bibinfo{person}{Meeyoung Cha}, {and} \bibinfo{person}{Kyungsin
  Park}.} \bibinfo{year}{2020}\natexlab{}.
\newblock \showarticletitle{Will Punishing Robots Become Imperative in the
  Future?}. In \bibinfo{booktitle}{\emph{Extended Abstracts of ACM CHI}}.
  \bibinfo{publisher}{Association for Computing Machinery}.
\newblock
\showISBNx{9781450368193}
\urldef\tempurl%
\url{https://doi.org/10.1145/3334480.3383006}
\showDOI{\tempurl}


\bibitem[\protect\citeauthoryear{Matthias}{Matthias}{2004}]%
        {matthias2004responsibility}
\bibfield{author}{\bibinfo{person}{Andreas Matthias}.}
  \bibinfo{year}{2004}\natexlab{}.
\newblock \showarticletitle{The responsibility gap: Ascribing responsibility
  for the actions of learning automata}.
\newblock \bibinfo{journal}{\emph{Ethics and Information Technology}}
  \bibinfo{volume}{6}, \bibinfo{number}{3} (\bibinfo{year}{2004}),
  \bibinfo{pages}{175--183}.
\newblock


\bibitem[\protect\citeauthoryear{Mirnig and Meschtscherjakov}{Mirnig and
  Meschtscherjakov}{2019}]%
        {mirnig2019trolled}
\bibfield{author}{\bibinfo{person}{Alexander~G Mirnig} {and}
  \bibinfo{person}{Alexander Meschtscherjakov}.}
  \bibinfo{year}{2019}\natexlab{}.
\newblock \showarticletitle{Trolled by the trolley problem: On what matters for
  ethical decision making in automated vehicles}. In
  \bibinfo{booktitle}{\emph{Proc. of ACM CHI}}. \bibinfo{pages}{1--10}.
\newblock


\bibitem[\protect\citeauthoryear{Mulligan}{Mulligan}{2017}]%
        {mulligan2017revenge}
\bibfield{author}{\bibinfo{person}{Christina Mulligan}.}
  \bibinfo{year}{2017}\natexlab{}.
\newblock \showarticletitle{Revenge against robots}.
\newblock \bibinfo{journal}{\emph{SCL Rev.}}  \bibinfo{volume}{69}
  (\bibinfo{year}{2017}), \bibinfo{pages}{579}.
\newblock


\bibitem[\protect\citeauthoryear{Nyholm}{Nyholm}{2018}]%
        {nyholm2018attributing}
\bibfield{author}{\bibinfo{person}{Sven Nyholm}.}
  \bibinfo{year}{2018}\natexlab{}.
\newblock \showarticletitle{Attributing agency to automated systems:
  Reflections on human--robot collaborations and responsibility-loci}.
\newblock \bibinfo{journal}{\emph{Science and Engineering Ethics}}
  \bibinfo{volume}{24}, \bibinfo{number}{4} (\bibinfo{year}{2018}),
  \bibinfo{pages}{1201--1219}.
\newblock


\bibitem[\protect\citeauthoryear{Nyholm and Smids}{Nyholm and Smids}{2016}]%
        {nyholm2016ethics}
\bibfield{author}{\bibinfo{person}{Sven Nyholm} {and} \bibinfo{person}{Jilles
  Smids}.} \bibinfo{year}{2016}\natexlab{}.
\newblock \showarticletitle{The ethics of accident-algorithms for self-driving
  cars: An applied trolley problem?}
\newblock \bibinfo{journal}{\emph{Ethical Theory and Moral Practice}}
  \bibinfo{volume}{19}, \bibinfo{number}{5} (\bibinfo{year}{2016}),
  \bibinfo{pages}{1275--1289}.
\newblock


\bibitem[\protect\citeauthoryear{Rahwan}{Rahwan}{2018}]%
        {rahwan2018society}
\bibfield{author}{\bibinfo{person}{Iyad Rahwan}.}
  \bibinfo{year}{2018}\natexlab{}.
\newblock \showarticletitle{Society-in-the-loop: programming the algorithmic
  social contract}.
\newblock \bibinfo{journal}{\emph{Ethics and Information Technology}}
  \bibinfo{volume}{20}, \bibinfo{number}{1} (\bibinfo{year}{2018}),
  \bibinfo{pages}{5--14}.
\newblock


\bibitem[\protect\citeauthoryear{Reddy, Cakici, and Ballestero}{Reddy
  et~al\mbox{.}}{2019}]%
        {reddy2019beyond}
\bibfield{author}{\bibinfo{person}{Elizabeth Reddy}, \bibinfo{person}{Baki
  Cakici}, {and} \bibinfo{person}{Andrea Ballestero}.}
  \bibinfo{year}{2019}\natexlab{}.
\newblock \showarticletitle{Beyond mystery: Putting algorithmic accountability
  in context}.
\newblock \bibinfo{journal}{\emph{Big Data \& Society}} \bibinfo{volume}{6},
  \bibinfo{number}{1} (\bibinfo{year}{2019}),
  \bibinfo{pages}{2053951719826856}.
\newblock


\bibitem[\protect\citeauthoryear{Rubel, Castro, and Pham}{Rubel
  et~al\mbox{.}}{2019}]%
        {rubel2019agency}
\bibfield{author}{\bibinfo{person}{Alan Rubel}, \bibinfo{person}{Clinton
  Castro}, {and} \bibinfo{person}{Adam Pham}.} \bibinfo{year}{2019}\natexlab{}.
\newblock \showarticletitle{Agency Laundering and Information Technologies}.
\newblock \bibinfo{journal}{\emph{Ethical Theory and Moral Practice}}
  \bibinfo{volume}{22}, \bibinfo{number}{4} (\bibinfo{year}{2019}),
  \bibinfo{pages}{1017--1041}.
\newblock


\bibitem[\protect\citeauthoryear{Saxena, Huang, DeFilippis, Radanovic, Parkes,
  and Liu}{Saxena et~al\mbox{.}}{2019}]%
        {saxena2019fairness}
\bibfield{author}{\bibinfo{person}{Nripsuta~Ani Saxena}, \bibinfo{person}{Karen
  Huang}, \bibinfo{person}{Evan DeFilippis}, \bibinfo{person}{Goran Radanovic},
  \bibinfo{person}{David~C Parkes}, {and} \bibinfo{person}{Yang Liu}.}
  \bibinfo{year}{2019}\natexlab{}.
\newblock \showarticletitle{How do fairness definitions fare? Examining public
  attitudes towards algorithmic definitions of fairness}. In
  \bibinfo{booktitle}{\emph{Proc. of the AAAI/ACM AIES}}.
  \bibinfo{pages}{99--106}.
\newblock


\bibitem[\protect\citeauthoryear{Schirmer}{Schirmer}{2020}]%
        {schirmer}
\bibfield{author}{\bibinfo{person}{Jan-Erik Schirmer}.}
  \bibinfo{year}{2020}\natexlab{}.
\newblock \bibinfo{booktitle}{\emph{Artificial Intelligence and Legal
  Personality: Introducing “Teilrechtsfähigkeit”: A Partial Legal Status
  Made in Germany}}.
\newblock \bibinfo{pages}{123--142}.
\newblock
\showISBNx{978-3-030-32360-8}
\urldef\tempurl%
\url{https://doi.org/10.1007/978-3-030-32361-5_6}
\showDOI{\tempurl}


\bibitem[\protect\citeauthoryear{Sparrow}{Sparrow}{2007}]%
        {sparrow2007killer}
\bibfield{author}{\bibinfo{person}{Robert Sparrow}.}
  \bibinfo{year}{2007}\natexlab{}.
\newblock \showarticletitle{Killer robots}.
\newblock \bibinfo{journal}{\emph{Journal of Applied Philosophy}}
  \bibinfo{volume}{24}, \bibinfo{number}{1} (\bibinfo{year}{2007}),
  \bibinfo{pages}{62--77}.
\newblock


\bibitem[\protect\citeauthoryear{Tsamados, Aggarwal, Cowls, Morley, Roberts,
  Taddeo, and Floridi}{Tsamados et~al\mbox{.}}{[n.d.]}]%
        {tsamadosethics}
\bibfield{author}{\bibinfo{person}{Andreas Tsamados}, \bibinfo{person}{Nikita
  Aggarwal}, \bibinfo{person}{Josh Cowls}, \bibinfo{person}{Jessica Morley},
  \bibinfo{person}{Huw Roberts}, \bibinfo{person}{Mariarosaria Taddeo}, {and}
  \bibinfo{person}{Luciano Floridi}.} \bibinfo{year}{[n.d.]}\natexlab{}.
\newblock \showarticletitle{The Ethics of Algorithms: Key Problems and
  Solutions}.
\newblock  (\bibinfo{year}{[n.\,d.]}).
\newblock


\bibitem[\protect\citeauthoryear{van~der Woerdt and Haselager}{van~der Woerdt
  and Haselager}{2019}]%
        {van2019robots}
\bibfield{author}{\bibinfo{person}{Sophie van~der Woerdt} {and}
  \bibinfo{person}{Pim Haselager}.} \bibinfo{year}{2019}\natexlab{}.
\newblock \showarticletitle{When robots appear to have a mind: The human
  perception of machine agency and responsibility}.
\newblock \bibinfo{journal}{\emph{New Ideas in Psychology}}
  \bibinfo{volume}{54} (\bibinfo{year}{2019}), \bibinfo{pages}{93--100}.
\newblock


\bibitem[\protect\citeauthoryear{Wallace}{Wallace}{1994}]%
        {wallace1994responsibility}
\bibfield{author}{\bibinfo{person}{R~Jay Wallace}.}
  \bibinfo{year}{1994}\natexlab{}.
\newblock \bibinfo{booktitle}{\emph{Responsibility and the moral sentiments}}.
\newblock \bibinfo{publisher}{Harvard University Press}.
\newblock


\bibitem[\protect\citeauthoryear{Whitman, Hsiang, and Roark}{Whitman
  et~al\mbox{.}}{2018}]%
        {whitman2018potential}
\bibfield{author}{\bibinfo{person}{Madisson Whitman}, \bibinfo{person}{Chien-yi
  Hsiang}, {and} \bibinfo{person}{Kendall Roark}.}
  \bibinfo{year}{2018}\natexlab{}.
\newblock \showarticletitle{Potential for participatory big data ethics and
  algorithm design: a scoping mapping review}. In
  \bibinfo{booktitle}{\emph{Proc. of the 15th Participatory Design Conference:
  Short Papers, Situated Actions, Workshops and Tutorial-Volume 2}}.
  \bibinfo{pages}{1--6}.
\newblock


\bibitem[\protect\citeauthoryear{Wong}{Wong}{2019}]%
        {wong2019democratizing}
\bibfield{author}{\bibinfo{person}{Pak-Hang Wong}.}
  \bibinfo{year}{2019}\natexlab{}.
\newblock \showarticletitle{Democratizing algorithmic fairness}.
\newblock \bibinfo{journal}{\emph{Philosophy \& Technology}}
  (\bibinfo{year}{2019}), \bibinfo{pages}{1--20}.
\newblock


\bibitem[\protect\citeauthoryear{Youn and Hall}{Youn and Hall}{2008}]%
        {youn2008gender}
\bibfield{author}{\bibinfo{person}{Seounmi Youn} {and}
  \bibinfo{person}{Kimberly Hall}.} \bibinfo{year}{2008}\natexlab{}.
\newblock \showarticletitle{Gender and online privacy among teens: Risk
  perception, privacy concerns, and protection behaviors}.
\newblock \bibinfo{journal}{\emph{Cyberpsychology \& Behavior}}
  \bibinfo{volume}{11}, \bibinfo{number}{6} (\bibinfo{year}{2008}),
  \bibinfo{pages}{763--765}.
\newblock


\bibitem[\protect\citeauthoryear{Zhu, Yu, Halfaker, and Terveen}{Zhu
  et~al\mbox{.}}{2018}]%
        {zhu2018value}
\bibfield{author}{\bibinfo{person}{Haiyi Zhu}, \bibinfo{person}{Bowen Yu},
  \bibinfo{person}{Aaron Halfaker}, {and} \bibinfo{person}{Loren Terveen}.}
  \bibinfo{year}{2018}\natexlab{}.
\newblock \showarticletitle{Value-Sensitive Algorithm Design: Method, Case
  Study, and Lessons}.
\newblock \bibinfo{journal}{\emph{Proc. ACM Hum.-Comput. Interact.}}
  \bibinfo{volume}{2}, \bibinfo{number}{CSCW}, Article \bibinfo{articleno}{194}
  (\bibinfo{year}{2018}).
\newblock
\urldef\tempurl%
\url{https://doi.org/10.1145/3274463}
\showDOI{\tempurl}


\end{thebibliography}
